\documentclass[a4paper,aps,nopacs,nokeys,twocolumn,twoside,superscriptaddress]{revtex4}

\usepackage{graphicx,epic,eepic,epsfig,amsmath,latexsym,amssymb,verbatim,revsymb}

\usepackage{color}
\usepackage{theorem}
\newtheorem{definition}{Definition}

\newtheorem{theorem}[definition]{Theorem}

\newcommand{\tr}{\operatorname{tr}}
\newcommand{\<}{\langle}
\renewcommand{\>}{\rangle}

\begin{document}

\title{Entanglement and separability of quantum harmonic oscillator systems at finite temperature} 
\author{Janet Anders\footnote{email: janet@qipc.org}}
\affiliation{Quantum Information Technology Lab, National University of Singapore, 2 Science Drive 3, Singapore 117542}
\author{Andreas Winter\footnote{a.j.winter@bris.ac.uk}}
\affiliation{Department of Mathematics, University of Bristol, University Walk, Bristol BS8 1TW, United Kingdom}
\affiliation{Quantum Information Technology Lab, National University of Singapore, 2 Science Drive 3, Singapore 117542}

\published{Quant. Inf. \& Comp. {\bf 8(3\&4)}:0245-0262 (2008)}

\maketitle


{\bf Abstract.} In the present paper we study the entanglement properties of thermal (a.k.a.~\emph{Gibbs}) states of quantum harmonic oscillator systems as functions of the Hamiltonian and the temperature. We prove the physical intuition that at sufficiently high temperatures the thermal state becomes fully separable and we deduce bounds on the critical temperature at which this happens. We show that the bound becomes tight for a wide class of Hamiltonians with sufficient translation symmetry.  We find, that at the crossover the thermal energy is of the order of the energy of the strongest normal mode of the system and quantify the degree of entanglement below the critical temperature. Finally, we discuss the example of a ring topology in detail and compare our results with previous work in an entanglement-phase diagram.

\section{Introduction} \label{sec:intro}
\noindent
Entanglement~\cite{EPR,Schroedinger} is one of the most characteristic features of quantum systems, and frequently the emergence of classicality is associated with the disappearance of entanglement from the state, i.e.~separable states~\cite{Werner} are deemed classical. This, and applications in quantum information theory, e.g.~\cite{BDSW,teleport,E91}, have motivated a large literature on the problem how to distinguish separable from entangled states, see e.g.~\cite{sep-review}. 

Here we are interested in the entanglement properties of thermal (Gibbs) states of certain simple quantum systems: the starting point is the physically intuitive observation that the ground state of a Hamiltonian of many interacting ``particles''  (we will adopt a more neutral, quantum information, language by calling them just ``sites'')  is usually entangled, and so are the thermal states at sufficiently low finite temperature~\cite{blah}. For most well-behaved systems one would also expect the complementary behaviour, namely that the thermal state at sufficiently high temperature loses all quantum character, and becomes separable, a concept
formalised in~\cite{Werner}. This is indeed true for finite systems of the spin lattice type, because their Hilbert space is finite dimensional, and in the limit of infinite temperature the thermal state converges to the maximally mixed state; then, results on the existence of a fully separable ball around the maximally mixed state can be invoked~\cite{sep-ball}, to conclude that at large temperature the thermal state is separable.

For infinite-dimensional or infinitely extended systems this type of argument is no longer available, yet the physical intuition remains the same. In particular, in the present paper we focus on systems with a finite number of coupled, infinite-dimensional harmonic oscillators. They provide a playground for testing ideas in quantum theory and quantum information. In particular, the class of quadratic Hamiltonians in all canonical coordinates $(x_1, p_1, x_2, p_2, \ldots) = (R_1,R_2,\ldots,R_{2n})$ provides a rich enough range of physical systems, yet the mathematics is simple enough so that almost everything can be understood exactly.

Our objective is the thermal state of such a system and the question how entangled it is for a given temperature, especially the determination of the critical temperature at which all entanglement vanishes from the state. Ultimately, we'd like to understand the continuum limit of this toy model: entanglement of quantum fields in the thermal state~\cite{Janet-etal,AV,HAV}, but the present simpler setting is interesting in itself, and will yield some decisive insights.

\medskip

Consider a system of $n$ quantum harmonic oscillators, interacting with a quadratic self-adjoint Hamiltonian, the most general form of which is 
\begin{equation} \label{eq:quadratic-H}
  H = \sum_{j,k=1}^{2n} c_{jk} R_j R_k + \sum_{j=1}^{2n} d_j R_j,
\end{equation}
with coefficients $c_{jk} = c_{kj}^*$ for the quadratic terms and displacement coefficients $d_j = d_j^*$ for the linear terms. This Hamiltonian might, for example, describe a system of interacting oscillators in a lattice with some internal interactions; for the moment we will however look at the most general case.  

As for spin systems, the thermal states at low temperature $T$,
\begin{equation} \label{eq:Gibbs}
  \rho_{\beta} = \frac{1}{Z} \, e^{-\beta H} \quad \mbox{with} \quad Z = \tr [e^{-\beta H}],
\end{equation}
are in general entangled~\cite{blah-harmonic}, where $\beta = (k_B T)^{-1}$ is the inverse temperature. Conversely, it is intuitively expected that for sufficiently high temperature the thermal states are separable. In this paper we prove this intuition mathematically, and give quantitative bounds on the \emph{critical temperature} at which separability kicks in, in Section~\ref{sec:highT}. We proceed to discuss the special case of a translation-invariant Hamiltonian (e.g.~a ring of coupled harmonic oscillators) in Section~\ref{sec:chain} for which the critical temperature bound is shown to be tight. We provide an in-depth discussion of these results when applied to the harmonic ring with nearest-neighbour couplings in Section~\ref{sec:chain-chain}. In Section~\ref{sec:e-measure} we give a brief discussion of a simple entanglement measure which can be evaluated efficiently for thermal states at all temperatures and relate it to an established measure, the (Gaussian) entanglement of formation. In the final Section~\ref{sec:discussion} we close with a discussion of the physical meaning of our results and a comparison to previous work.

\section{Continuous Variable States}\label{sec:normal}
\noindent
Quadratic Hamiltonians are the working horse of theoretical physics and many a theory is based on the understanding of these simplest of models.  So it is in the case of continuous variable states, which have been studied extensively on the model of Gaussian states (for reference, see for instance~\cite{xxx,Braunstein:vanLoock}). Gaussian states can in particular be represented as the thermal states of a system with a quadratic Hamiltonian (\ref{eq:quadratic-H}) and we can therefore use the well-developed machinery of analysing Gaussian continuous variable states for our purpose.  In particular, the thermal states under discussion are uniquely specified by the expectation values of all position and momentum variables, i.e. $\< R_j \>$ which vanish for all $j$,  and by the quadratures (covariances) 
$\< \frac{1}{2}\{R_j, R_k\} \> = \< \frac{1}{2}(R_j R_k + R_k R_j) \>  \equiv \gamma_{jk}$. 
The latter form a $2n \times 2n$-matrix, $\gamma$, which is real symmetric and satisfies the so-called \emph{uncertainty relations} $\gamma + i \frac{\hbar}{2} \sigma \geq 0$, with the symplectic form $\sigma$ determined through the commutation relations $[R_j, R_k] = i \hbar \, \sigma_{jk}$ and given explicitly by 
\begin{equation}
  \sigma = \bigoplus_{j=1}^n \left[ \begin{array}{rr}
                                      0 & 1 \\
                                     -1 & 0
                                    \end{array} \right].
\end{equation}
These uncertainty relations are sufficient for a covariance matrix (CM) to define a bona-fide quantum state and the first moments can be changed arbitrarily by applying local, i.e.~single-oscillator,  Weyl-displacement operators. Therefore, only the quadratic part of the Hamiltonian in  Eq.~(\ref{eq:quadratic-H}) is relevant and the coefficients $d_j$ can be set to zero. Thus the entanglement properties of $\rho_{\beta}$ are in general determined by the covariance matrix $\gamma$ and also higher moments; in the Gaussian case by the covariance matrix alone. 

\medskip

\begin{figure} [htbp]
\centerline{\epsfig{file=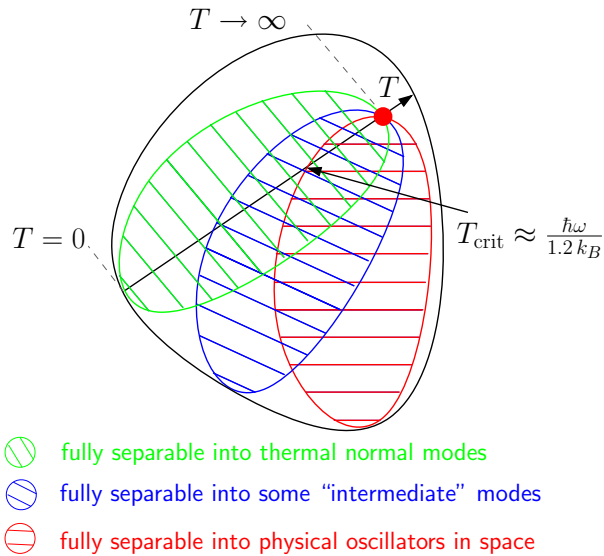, width=8cm}} 
\vspace*{13pt}
\caption{\label{fig:coordinates} Separability is a concept depending on the choice of the basis of the system. Shown here is the set of all states (black set) which a system of $n$ harmonic oscillators can realise in principle. We wish to look at only the thermal states for a given Hamiltonian and vary the temperature $T$ (black line).
  On a journey from high temperatures towards low temperatures we pass through the following stages: At extremely high temperatures we expect that the system is fully separable for all possible decompositions into $n$ modes. This is because high temperature provides so much energy that the oscillators lose all correlations and behave completely independent from each other in \emph{any} representation.
When the temperature is lowered the system reaches a point where \emph{something} in the chain  must become entangled. The real space representation referring to the actual lattice sites turns out to be the representation where entanglement can be found first, at least for Hamiltonians with translational invariance of the coupling
(as discussed in Section~\ref{sec:chain}) which implies that the derived bound for the critical temperature $T_{\rm {crit}}$, see Eq.~(\ref{eq:energy-scale}) in Section~\ref{sec:chain}, is in fact tight. In other words, for temperatures below $T_{\rm {crit}}$ the thermal state picks up entanglement between sites in space and leaves the set of fully separable sites (red). 
However, any ``intermediate'' mode representation between physical-space and the normal mode decomposition will pick up entanglement only at smaller and smaller temperatures (blue). The extreme case is of course the normal mode basis for which no temperature is low enough to produce entanglement and any thermal state remains separable (green).}
\end{figure}

\textbf{Remark~1}
Even though the following will be obvious to some readers, let us here make a remark regarding the choice of the basis for the class of systems under discussion to avoid a common confusion. The original physical oscillators described by the canonical observables $x_j = R_{2j-1}$ and $p_j = R_{2j}$ may disappear from sight in an alternative mathematical description of the system. In fact, any symplectic transformation $S$ transforms into a collection of $n$ abstract modes, $R'^T = S R^T$, that will be coupled in a different way from the original oscillators. This change of coupling results in a change of the entanglement properties between the underlying modes; for instance the normal modes are by definition completely decoupled and therefore \emph{never} entangled at any temperature, see Fig.~\ref{fig:coordinates}.
It is hence useless to speak about complete separability in an abstract way. There is no such thing -- separability may be had in one basis or decomposition but not in a rotated, different basis at the same temperature, depending on the choice of the division. It is not only a grouping issue of how many oscillators one puts  in several groups -- the issue occurs already when the choice of the representation is made -- real space (physical oscillators) or k-space (normal modes or phononic modes, compare \cite{Janet}) or any ``intermediate'' set of modes. 

\section{High Temperature implies Separability} \label{sec:highT}
\noindent
We will show that for sufficiently large $T$, i.e.~sufficiently small $\beta$, the thermal state is separable in the sense that it is a mixture of single-oscillator Gaussian states.  This holds true for any arbitrary choice of basis of the system, i.e.  separable with respect to the original spatial basis of physical oscillators ($R$) or w.r.t. the normal basis of decoupled oscillators ($R'$) or any other basis that can be reached by symplectic transformations.

Separability of a Gaussian state can be formulated in terms of the covariance matrix alone. Werner and Wolf \cite{Werner:Wolf} have shown that a bipartite Gaussian state, with its covariance matrix denoted $\gamma$, is separable if and only if there exist Gaussian states on the local systems with covariance matrices $\eta_A$ and $\eta_B$ such that
\begin{equation} \label{eq:gamma-sep}
	\gamma \geq \eta_A \oplus \eta_B.
\end{equation} 
Note here, that matrix inequalities for matrices containing physical units, such as position and momentum, should be read as being evaluated with respect to vectors $| v \>$ which themselves carry units; for instance $\gamma \geq \eta_A \oplus \eta_B$ shall be read 
as $\< v| \gamma| v \>  \geq \< v| \eta_A \oplus \eta_B | v \> $ for all vectors $| v\>$.

In easy generalisation of Eq.~(\ref{eq:gamma-sep}), our thermal state $\rho_{\beta}$ of $n$ oscillators with covariance matrix $\gamma_{\beta}$ is fully separable into single physical oscillators if and only if there exist Gaussian states of the individual oscillators, with covariance matrices $\eta_j$ ($j=1,\ldots,n$) such that 
\begin{equation} \label{eq:multi-sep}
	\gamma_{\beta} \geq \bigoplus_{j=1}^n \eta_j. 
\end{equation}
We shall show that for any arbitrary $n$-tuple of one-mode covariance matrices 
$\eta_j$ (each a $2 \times  2$ matrix) this eventually becomes true for high 
enough temperatures. The reasoning is essentially as follows: The matrix 
$\bigoplus_{j=1}^n \eta_j$ has a fixed largest eigenvalue, denoted $\lambda_{\max}$. 
However the \emph{smallest} eigenvalue of $\gamma_{\beta}$ turns out to diverge 
as $\beta \to 0$ and therefore the matrix inequality must hold above some finite 
critical temperature $T_{\rm {crit}}$.

\medskip

\begin{theorem} \label{thm:highT}
	For every quadratic Hamiltonian of the form~(\ref{eq:quadratic-H}), 
	there exists a temperature $T_{\rm {crit}}$ 
	such that for all $T \geq T_{\rm {crit}}$ 
	the thermal state $\rho_{\beta}$ with $\beta = (k_B \, T)^{-1}$ is fully separable.
\end{theorem}

\medskip

{\bf Proof:} There exists a symplectic transformation matrix $S$ of the canonical variables $R'^{\,T} = S R^T$, which diagonalises the Hamiltonian (\ref{eq:quadratic-H}) with respect to the primed variables,
\begin{equation}
    	\label{eq:diagonalH}
    	H = \sum_{j=1}^{n} H_j = \sum_{j=1}^{n} \frac{1}{2} 
	\left(\alpha_{2j- 1}\, R_{2j -1}^{' 2} + \alpha_{2j} \, R_{2j}^{' 2}  \right),
\end{equation}  
where $S$ is chosen such that $\sum_{m,n} \, S^{-T}_{km} \, c_{mn} \, S^{-1}_{nl} = \alpha_l \, \delta_{k,l}$. Any thermal state $\rho_{\beta}$ can thus be expressed as a (tensor) product of Gaussian states of single harmonic modes in the corresponding normal basis,
\begin{equation}
	\rho_{\beta} \propto \bigotimes_{j=1}^n e^{-\beta H_j}.
\end{equation}
The condition that the new variables are canonical is expressed by the symplectic condition $S \, \sigma \, S^T = \sigma$ and the covariance matrix in these new coordinates is simply $\gamma_{\beta}' = S \, \gamma_{\beta} \, S^T$.  Since $S$ is invertible, the separability condition (\ref{eq:multi-sep}) can be brought over to the primed matrices. The reason is that even though the symplectic transformation $S$ may change the eigenvalues of a positive matrix it will not change its positivity. Therefore the separability condition (\ref{eq:multi-sep}) becomes 
\begin{equation} \label{eq:sep-cond}
    \gamma_{\beta}' =    S \, \gamma_{\beta} \, S^T 
            \geq S \left( \bigoplus_{j=1}^n \eta_j \right) S^T
            =:   \Gamma_0,
\end{equation}
for some constant covariance matrix $\Gamma_0$ with largest eigenvalue $\lambda_0$. On the left hand side, however, we have the covariance matrix of a tensor product state,  i.e. the covariance matrix $\gamma_{\beta}'$  is the direct sum of $2\times 2$-blocks, 
\begin{equation}
	\gamma_{\beta}' = \bigoplus_{j=1}^n \gamma'_j (\beta),
\end{equation}
where $\gamma'_j (\beta)$ are the covariance matrices of the thermal states of the harmonic modes each with their Hamiltonians $H_j$, at inverse temperature $\beta$. These blocks are given explicitly by 
\begin{equation} \label{eq:blockgammas}
  	\gamma'_j (\beta) = \frac{\hbar}{2} \coth \frac{\beta \hbar \omega_j}{2}
		 \left[ \begin{array}{cc} 
			\frac{1}{m_j \omega_j} \, & 0 \\ 
			0 & m_j \omega_j 		
		\end{array} \right],  
  \end{equation}
where $\omega_j = \sqrt{\alpha_{2j-1} \, \alpha_{2j}}$ is the frequency of each of the modes and $m_j = 1/ \alpha_{2j}$ is the mass of the physical oscillators (see~\cite{Anders-dipl}  for reference). 

\medskip

Introducing the length unit $L_j$ we rescale the variance in position
$ A_j =\frac{1}{L_j^2} \,  \frac{\hbar}{2 m_j \omega_j}$, and with $\hbar/L_j$ being the appropriate fundamental momentum unit the variance in momentum becomes
$ B_j = \frac{\hbar m_j \omega_j}{2} \frac{L_j^2}{\hbar^2}$.
The final unit-free covariance matrices for each independent normal mode are then 
\begin{equation}
	\tilde \gamma'_j (\beta) = 
	\left[ \begin{array}{cc} 
		A_j \,\coth \frac{\beta \hbar \omega_j}{2}  & 0 \\ 
		0& B_j \, \coth \frac{\beta \hbar \omega_j}{2}  
	\end{array} \right],
\end{equation} 
with the unit-free eigenvalues on the diagonal which indeed go to infinity for $\beta \to 0$ for any arbitrary finite unit-length $L_j$. 
Alternatively, one can calculate the symplectic eigenvalues \cite{footnote1} of the unit-free matrices as
$\tilde s_j = \sqrt{\det \tilde \gamma'_j (\beta)}
            = \frac{1}{2} \coth \frac{\beta \hbar \omega_j}{2}$, so that
$\tilde s_j \ge \frac{1}{2}$ for all  temperatures and $\tilde s_j \to \infty$ for $\beta \to 0$ as well. 
In other words, for suffiently small $\beta$ (i.e., sufficiently large
$T$), the matrix inequality (\ref{eq:multi-sep}) will be satisfied,
proving the thermal state separable. $\square$\,.

\medskip

So far we could confirm the intuition that for $\beta \to 0$ the thermal states $\rho_{\beta}$  take on a separable configuration for any Hamiltonian and irrespective of the chosen mode-decomposition. However, we want to improve our argument such that we can predict the quantitative critical temperature for a particular (class of) Hamiltonian at which the thermal state undergoes the transition from entangled to separable. For this purpose, let us go through the above proof once more, but now paying attention to the freedoms we have there to get as small as possible a bound on the critical temperature -- thus turning the quest for the exact value into an optimisation problem.

\medskip

Let $S$ be again the symplectic transformation that diagonalises $\gamma_{\beta}$, and rescale to the unit-free version of the matrix,
\begin{equation}\label{eq:William}
	\gamma_{\beta} \overset{S}{\mapsto} \gamma_{\beta}' 
	\overset{\rm units}{\longmapsto} \tilde \gamma_{\beta}' 
	= \bigoplus_{j=1}^n \left[\begin{array}{cc} 
		\tilde s_j &0 \\ 
		0 & \tilde  
	s_j \end{array} \right].
\end{equation} 
Now consider all unit-free covariance matrices $\tilde \Gamma_0$ that can be reached by first choosing individual covariance matrices for all modes [the $\eta_j$ for $j=1,\ldots, n$ in Eq.~(\ref{eq:multi-sep})], then, secondly, transforming them with $S$, $\Gamma_0 = S\left( \bigoplus_{j=1}^n \eta_j \right) S^T$, and finally rescaling to the unit-free matrix $ \tilde \Gamma_0$. We require that every eigenvalue of $\tilde \gamma_{\beta}'$ [see Eq.~(\ref{eq:William})] must be larger than every eigenvalue of $\tilde \Gamma_0$, in particular larger than the maximal eigenvalue $\lambda_0 = \lambda_{\max} (\tilde \Gamma_0)$, 
\begin{equation}
	\tilde s_j = \frac{1}{2} \coth \frac{\beta \hbar \omega_{j}}{2} \ge  \lambda_0 \quad 
	 \mbox{ for all } j = 1, \ldots, n.
\end{equation}
Therefore the separability condition for the thermal states $\rho_{\beta}$ is sufficiently satisfied when $\beta$ obeys 
\begin{equation} \label{eq:beta-cond}
  	\beta_{\rm {crit}}  \le \frac{1}{\hbar \omega_{\max}} 
	\ln\left(\frac{2 \lambda_0 +1}{2 \lambda_0 -1} \right).
\end{equation}
This still is a very rough bound for the critical temperature, $T_{\rm {crit}}  = \frac{1}{k_B \beta_{\rm {crit}}}$, as there is no reason in general to expect that we have to compare the smallest eigenvalue on the left with the largest on the right, i.e. this condition may not be necessary for separability. Nevertheless, a slight refinement of this approach can give reasonable and intuitively appealing bounds on the critical temperature, which are in fact sometimes \emph{exact}, as we shall show in the next section.

\section{Symmetric systems} \label{sec:chain}

The recipe of the previous section turns out to be pretty good for systems with a lot of symmetry, and some additional structure of the Hamiltonian: Following Audenaert \textit{et al.}~\cite{AEPW}, we consider systems with a Hamiltonian of the form (now in position and momentum coordinates)
\begin{equation} \label{eq:kinetic+potential}
  H = \sum_j \frac{1}{2m} p_j^2 + \sum_{jk} \frac{m}{2}V_{jk}x_j x_k,
\end{equation}
with $V_{jk} = V_{kj} = V_{\pi(j) \, \pi(k)}$ for any permutation $\pi$.
(A particular such system is the translation invariant ring of $n$ equal harmonic oscillators, coupled via a nearest-neighbour interaction $\omega$ and exposed to on-site potential traps with strength $\delta$, see Section \ref{sec:chain-chain}.)

Note, that the $x_j$ all commute, hence we may assume that the \emph{potential matrix} $\frac{1}{2}V$ is real symmetric and for physical reasons it needs to be positive semi-definite so that the energy is bounded from below. Then it is shown in~\cite{AEPW} that the covariance matrix of the thermal state at inverse temperature $\beta$ can be written as a direct sum of a position part, $\gamma_x (\beta)$, and a momentum part, $\gamma_p (\beta)$, 
$\gamma_{\beta} = \gamma_x(\beta) \underset{xp}{\oplus} \gamma_p(\beta)$, with
\begin{equation}  \label{eq:CV}
	\begin{split}  
		\gamma_x(\beta) = \frac{\hbar}{2m} \,  V^{-1/2} \coth \frac{\beta \hbar V^{1/2}}{2},   \\
  		\gamma_p(\beta) = \frac{m\hbar}{2} \,  V^{ 1/2\phantom{-}} \coth \frac{\beta \hbar V^{1/2}}{2}.
	\end{split}
\end{equation}
Writing the covariance matrix in this way requires reordering of the canonical variables into all the positions first followed by all their corresponding momenta. The observation in~\cite{AEPW} was, that because of the block nature of $\gamma_{\beta}$ it can actually be diagonalised by an \emph{orthogonal} symplectic transformation of the special form $S \underset{xp}{\oplus} S$, with an orthogonal $n\times n$-matrix $S$. As a consequence of symplecticity, $S \underset{xp}{\oplus} S$ decouples the oscillators in the sense that in the new canonical coordinates $x_j' = \sum_k S_{jk} x_k$ and $p_j' = \sum_k S_{jk} p_k$, the Hamiltonian assumes the form
\begin{equation}
  H = \sum_j \frac{1}{2m}{p_j'}^2 + \sum_j \frac{m}{2}\omega_j^2 {x_j'}^2.
\end{equation}
In the canonical coordinates $x_j'$, $p_j'$ the covariance matrix $\gamma'_{\beta}$ becomes a direct sum of blocks (compare \eqref{eq:blockgammas})
\begin{equation}
  \gamma_j' (\beta)= \left[ \begin{array}{cc}
           \frac{\hbar}{2m\omega_j} \coth\frac{\beta\hbar\omega_j}{2} & 0 \\
           0 &  \frac{m\hbar\omega_j}{2} \coth\frac{\beta\hbar\omega_j}{2}
         \end{array} \right],
\end{equation}
where the masses are all the same and the frequencies $\omega_j$ are obtained by diagonalising
$\sqrt{V}$; indeed, the $\omega_j^2$ are just the eigenvalues of the real symmetric
matrix $V$.

We intend to compare $\gamma_{\beta}$ with $\bigoplus_{j=1}^n \eta_0$, a direct sum of $2\times 2$-blocks $\eta_0$, which we argue can be chosen w.l.o.g.~as
\begin{equation}
  \eta_0 = \left[ \begin{array}{cc}
                      \frac{\hbar}{2m\omega_0} & 0 \\
                      0 & \frac{m\hbar\omega_0}{2}
                    \end{array} \right].
\end{equation}
This is straightforward since first, the matrix ordering is preserved under the map that sets all ``mixed term'' covariances to zero. Secondly, the state can be assumed as a minimal uncertainty state -- otherwise we could decrease either of the diagonal entries of $\eta_0$. As a consequence of orthogonality of $S \underset{xp}{\oplus} S$, $\bigoplus_{j=1}^n \eta_0$ is
of the exact same form in the transformed coordinates $x_j'$, $p_j'$:
\begin{equation}
  \left(S \underset{xp}{\oplus} S \right) \, 
  \bigoplus_{j=1}^n \eta_0 \, 
  \left(S \underset{xp}{\oplus}  S \right)^T 
  = \bigoplus_{j=1}^n \eta'_0,
\end{equation}
with $\eta'_0$ having a possibly re-scaled frequency $\omega'_0$, but since $\omega_0$ is still arbitrary we drop the prime immediately.

\medskip

\begin{figure} [htbp]
\centerline{\epsfig{file=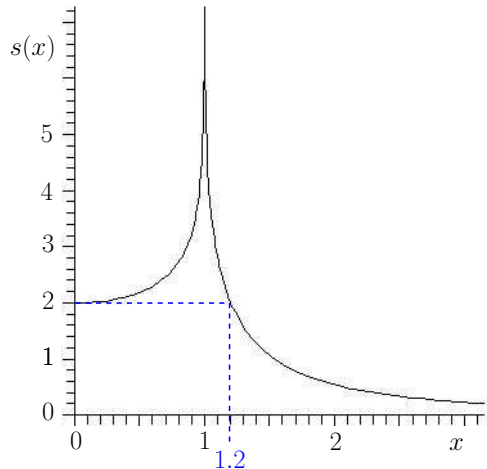, width=7cm}} 
\vspace*{13pt}
\caption{\label{fig:s} The scaling function $s(x)$ is monotonically increasing left of the singularity at $x=1$ and decreasing right of it. This function encodes the multiplicative offset from $\frac{1}{k_B}\hbar\omega_{0}$ for the critical temperature. The dashed blue line indicates the cutting point $x = 1.199678 \approx 1.2$ where $s(x)$ passes below its initial value of 2.}
\end{figure}

This implies that we get $2n$ inequalities which have to be satisfied by $\beta$ to guarantee separability: for all $j$,
\begin{equation} 
  \coth\frac{\beta\hbar\omega_j}{2} \geq \frac{\omega_j}{\omega_0} 
  \quad \mbox{and} \quad \coth\frac{\beta\hbar\omega_j}{2} \geq \frac{\omega_0}{\omega_j}.
\end{equation}
(Note, that by $\coth x \geq 1$ only one of the two inequalities expresses a nontrivial constraint for every $j$, so we have effectively only $n$ inequalities.)
A brief calculation (distinguishing the cases of $\omega_j \leq \omega_0$ and $\omega_j \geq \omega_0$) shows these constraints require
\begin{equation}
  \beta \leq \min_j \left\{ \frac{1}{\hbar\omega_j}
                      \ln \left| \frac{\omega_j+\omega_0}{\omega_j-\omega_0} \right| \right\}.
\end{equation}
Introducing the \emph{scaling function}
\begin{equation}
  s(x) := \frac{1}{x}\ln\left| \frac{1+x}{1-x} \right|,
\end{equation}
we can rephrase this as
\begin{equation}
  \beta \leq \frac{1}{\hbar\omega_0} \min_j s\left(\frac{\omega_j}{\omega_0}\right).
\end{equation}
The good thing here is that we see how the critical temperature depends only on a few universal parameters and items: essentially, $\beta$ is given by $\frac{1}{\hbar\omega_0}$, with the scaling function giving a constant multiple. The properties of $s$ are very simple: it is monotonically increasing in the interval $[0;1)$, and decreasing in the interval $(1;\infty)$, and $s(x)\geq 2$ for $0\leq x<1$ (see Fig.~\ref{fig:s}). 
Indeed, the Taylor expansion of $s$ at $x=0$ is
\begin{equation}
  \label{eq:s-taylor}
  s(x) = 2\sum_{k=0}^\infty \frac{x^{2k}}{2k+1},
\end{equation}
and it obeys the functional relation
\begin{equation}
  \label{eq:s-functional}
  s(x) = \frac{1}{x^2} s\left( \frac{1}{x} \right).
\end{equation}

\medskip

This implies that actually only the extremal values out of the eigenfrequency spectrum of $H$, $\{\omega_j\}$, become relevant:
\begin{equation} \label{eq:beta-propto-spectrum}
  \beta(\omega_0) = \frac{1}{\hbar\omega_0} 
                    \min \left\{ s\left(\frac{\omega_{\min}}{\omega_0}\right), 
                                 s\left(\frac{\omega_{\max}}{\omega_0}\right) \right\},
\end{equation}
where now $\beta(\omega_0)$ is an upper bound so that $\beta \le \beta(\omega_0)$ implies separability.
Because we want to use this to find (or bound) the minimum temperature at which
the thermal state is separable, we would like to maximise $\beta(\omega_0)$
by choosing $\omega_0$ appropriately. 
For example, we find a physically intuitive bound for the critical temperature at which all entanglement vanishes from the oscillator system: by choosing $\omega_0=\omega_{\max}$, we find
$\beta(\omega_{\max}) = \frac{1}{\hbar\omega_{\max}} s\left(\frac{\omega_{\min}}{\omega_{\max}}\right)$, with values of $s$ only in the interval $[0;1]$ where they are always larger or equal to 2. Hence, if the thermal energy available to the system, $k_B T$, is at least half as big as the largest energy of an eigenmode of the system, $\hbar\omega_{\max}$, the system will be separable.

\medskip

By a simple variational argument it is seen that to maximise $\beta(\omega_0)$ it is necessary that $s\left(\frac{\omega_{\min}}{\omega_0}\right) = s\left(\frac{\omega_{\max}}{\omega_0}\right)$, and since this has a unique solution for $\omega_0$, this value is indeed the one attaining the maximum; i.e., if $\beta \leq \beta_{\rm {crit}} := \max_{\omega_0} \beta(\omega_0)$, then the thermal state at inverse temperature $\beta$ is fully separable.
In fact, the solution to the equation for $s$ depends only on the ratio of the spectrum, $r := \frac{\omega_{\max}}{\omega_{\min}} \ge 1$, and we can summarise our findings so far in the following theorem.

\medskip

\begin{theorem}
  \label{thm:beta_crit}
  For all inverse temperatures 
  $\beta \leq \beta_{\rm{crit}} := \frac{1}{\hbar\omega_{\max}} \sigma(r)$,
  the thermal state is separable. Here, $\sigma(r)=t s(t)$ with the unique
  $1\leq t\leq r$ such that $s(t) = s\left( \frac{t}{r} \right)$.
\end{theorem}

\medskip

{\bf Proof:} We simply go with $\omega_0 = \frac{1}{t}\omega_{\max}$ into the above  argument; there, it is straightforward to see that nothing is gained by choosing $\omega_0$ outside the interval $[\omega_{\min};\omega_{\max}]$ $\square$\,.

\begin{figure} [htbp]
\centerline{\epsfig{file=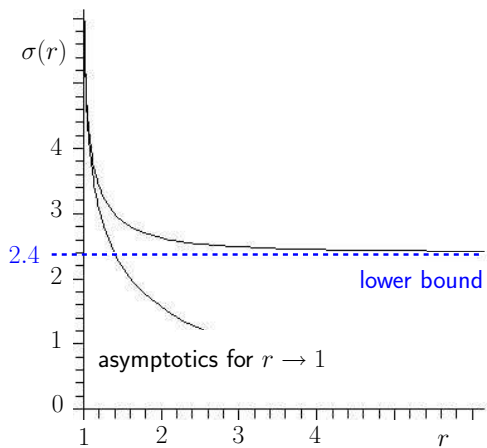, width=7cm}} 
\vspace*{13pt}
\caption{\label{fig:sigma} A plot of $\sigma$ as a function of $r$ comparing it with its two asymptotic lower bounds: The constant $399357\ldots \approx 2.4$ in the limit $r \to \infty$ (see also Fig.~\ref{fig:s}) and the expression from Eq.~\eqref{eq:sigma-asy} for $r \to 1$.}
\end{figure}

\medskip
By looking at the graph of $s$ once more (Fig.~\ref{fig:s}), we find that $\sigma(r)$ is monotonically decreasing with $r$. Its infimum is attained at $r=+\infty$, where $t>1$ is the unique solution to the equation $s(t)=2$, yielding $t = 1.199678\ldots$; hence $\sigma(+\infty) = 2\cdot 1.199678\ldots = 2.399357\ldots$. On the other hand, for very small $r$ (i.e., close to $1$), we can equally easily understand the asymptotics of $\sigma(r)$: by choosing $t=\sqrt{r}$ -- a good choice as can be seen by looking at the functional equation~(\ref{eq:s-functional}) -- we find 
\begin{equation} \label{eq:sigma-asy}
	\begin{split}
		\sigma(r) \geq \min \left\{ s\left(\sqrt{r}\right), s\left(\frac{1}{\sqrt{r}}\right) = r\,s\left(\sqrt{r}\right) \right\} \\
		=    s\left(\sqrt{r}\right) \geq \ln \frac{2}{\sqrt{r}-1} \geq \ln \frac{4}{\epsilon},
	\end{split}
\end{equation}
for $r = 1 + \epsilon$. Indeed this is asymptotically exact as $r \rightarrow 1+0$;
see Fig.~\ref{fig:sigma}.

\medskip

While all this might feel quite ad hoc (though giving a reasonable bound), we want to show now that the described method to obtain $\beta_{\rm{crit}}  $ actually yields the \emph{exact} cutoff point for all systems of the type (\ref{eq:kinetic+potential}) with sufficient translation symmetry. To be precise, 
let $G$ be the group of site permutations which leave the Hamiltonian invariant. Then
we have the following complement to Theorem~\ref{thm:beta_crit}:

\medskip

\begin{theorem}
  \label{thm:really-crit}
  Under the previous assumptions on the Hamiltonian, and if $G$ acts
  transitively on the sites of our $n$ oscillators (meaning that for every
  two sites $j$ and $k$ there is a permutation $\pi\in G$ such that
  $\pi(j)=k$), then the thermal state at inverse
  temperature $\beta > \beta_{\rm{crit}}$ is \emph{not} fully 
  separable into physical oscillators.
\end{theorem}

\medskip

{\bf Proof:} For this, all we have to show is that the ansatz of comparing $\gamma$ to $\bigoplus_{j=1}^n \eta_0$ is already the most general we need to consider. Indeed, the covariance matrix $\gamma$ inherits the symmetry of $H$, now on the level of pairs of canonical coordinates; but then, if $\gamma \geq \bigoplus_{j=1}^n \eta_j$, also $\gamma \geq \bigoplus_{j=1}^n \eta_{\pi(j)}$ for any permutation $\pi\in G$, and since $G$ acts transitively, we find by averaging that $\gamma \geq \bigoplus_{j=1}^n \eta_0$, with
\begin{equation}
	\eta_0 = \frac{1}{|G|}\sum_{\pi\in G} \eta_{\pi(j)}
             	= \frac{1}{n}\sum_{j=1}^n \eta_j.
\end{equation}
Hence, all of our calculations above were indeed w.l.o.g.~and there is no way we could  have found a smaller critical temperature $\square$\,.

\medskip

Examples of transitively acting symmetry group of the Hamiltonian abound: for instance the shift-invariant harmonic ring considered in the following Section~\ref{sec:chain-chain}, or any other periodic lattice in higher dimension.

\medskip

\begin{figure} [htbp]
\centerline{\epsfig{file=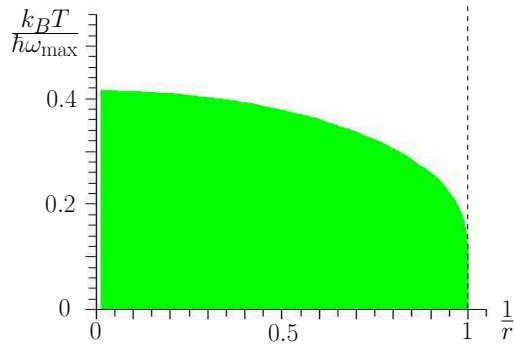, width=7cm}} 
\vspace*{13pt}
\caption{\label{fig:universal-phase} This is the universal entanglement \emph{phase diagram} of translation symmetric systems of type Eq.~(\ref{eq:kinetic+potential}): The boundary marks the line between the system being fully separable into separate sites for high temperatures, and it being (somehow) entangled for low temperatures (green shaded).    The temperature on the vertical axis is scaled with $\hbar\omega_{\max}/k_B$, the horizontal axis is the inverse of the spectral ratio, $\frac{1}{r} = \frac{\omega_{\min}}{\omega_{\max}}$.}
\end{figure}
This result shows that the critical temperature for the class
of systems considered in the theorem depends really only on the top
end of the normal mode spectrum $\omega_{\max}$,
and the ratio $r$ of the spectrum; in Fig.~\ref{fig:universal-phase}
we interpret this result as describing a universal phase diagram
of the presence of entanglement in the system.

\medskip

  \textbf{Remark~2}
  Here we point out why all the assumptions of Theorem~\ref{thm:really-crit}
  are necessary in our proof: we have to impose translation symmetry to be able to reduce
  to the case of checking the separability criterion on a direct sum of
  identical blocks $\eta_0$. 
  Then we need the assumption of type (\ref{eq:kinetic+potential}) 
  Hamiltonian for two steps: the first is to be able to  get rid of the
  off-diagonal entropies of $\eta_0$, so that $\bigoplus_{j=1}^n \eta_0$
  is the direct sum of two multiples of the identity (collecting
  the position and momentum variables in contiguous blocks); the second
  is to have an orthogonal symplectic transformation of the form $S \underset{xp}{\oplus} S$
  diagonalising $H$, which leaves $\bigoplus_{j=1}^n \eta_0$ alone.

\section{The Harmonic Ring} \label{sec:chain-chain}
\noindent
Let us take a particularly symmetric system, the translation invariant ring of $n$ equal harmonic oscillators, coupled via a nearest-neighbour interaction $\omega$ and exposed 
to on-site potential traps with strength $\delta$. The Hamiltonian of such
a quantum system is
\begin{equation} \label{eq:hamiltonian-ring}
 	H = \sum_{j=1}^n \left( \frac{p_j^2}{2m} 
		+ \frac{m \delta x_j^2}{2} + \frac{m \omega (x_{j+1} - x_j)^2}{2} \right),
\end{equation}
where $m$ is the mass of each physical oscillator. (The convention is that indices
are understood $\rm {mod} n$.)
Diagonalising via discrete Fourier-transformation leads to a
Hamiltonian with $n$ decoupled oscillators with the eigenfrequencies 
\begin{equation} \label{eq:dispersion}
	\omega_{j} =  2 \omega \sqrt{\sin^2 \frac{\pi j}{n}  
     			+ \left( \frac{\delta}{2 \omega} \right)^2} \rm { with } j = 0, \ldots,  n-1.
\end{equation}
These eigenfrequencies are on the scale of the interaction frequency $\omega$
but depend heavily  on the trapping potential $\delta$ as well.  In the special
case $\delta = 0$, a particular spatial symmetry is assumed, i.e. the coordinates 
of the centre of mass of all oscillators remains completely free. However, for 
any finite, ever so tiny,  $\delta$ this degree of freedom is harmonic. 
The scale of energy in this discussion is given by the gauge frequency
$\omega_0$ where $ \omega_{\min} \le \omega_0 \le \omega_{\max}$ can be chosen 
in an optimal way to achieve the tightest bound. Roughly, $\omega_0$ will 
be $\omega _{\max}$, and the correction to that depends only on the ratio 
of the spectrum $r = \frac{\omega_{\max}}{\omega_{\min}}$ which itself depends on $\delta$.

According to Eq.~(\ref{eq:dispersion}) the minimum frequency which gives 
the cost of the lowest oscillation is $\omega_{\min} =  \delta$ and the maximum 
frequency is 
$\omega_{\max} =  2 \omega \sqrt{1 + \left( \frac{\delta}{2 \omega} \right)^2}$. 
The ratio between these two is then 
$r =  \sqrt{1 + \left( \frac{2 \omega}{\delta} \right)^2}$ and the 
critical temperature becomes, by Theorem \ref{thm:really-crit},
\begin{equation}
	k_B T_{\rm {crit}} = \frac{\hbar \omega_{\max}}{\sigma(r)}
                        = \hbar \delta \frac{r}{\sigma\left( r\right)},
\end{equation}
which is an implicit function of $\frac{\delta}{\omega}$ via
$r = r\left(\frac{\delta}{\omega}\right)$.
For $\delta \to 0$, the ratio is approximately
$r \approx \frac{2 \omega}{\delta} \gg 1$, and therefore 
\begin{equation} \label{eq:T_crit-low}
	\frac{k_B T_{\rm {crit}}}{\hbar \omega}
     	\approx  \frac{2}{\sigma\left(2 \frac{\omega}{\delta}\right) } 
	    \approx  \frac{1}{1.2},
\end{equation}
where $\sigma\left(2 \frac{\omega}{\delta}\right) \approx 2 \cdot 1.2$ for small $\delta$. 
In the limit where $\delta \to \infty$, the ratio becomes $r \sim 1 + \frac{2 \omega^2}{\delta^2}$ and $\sigma(r) \approx \ln \frac{2 \delta^2}{\omega^2}$ and therefore 
\begin{equation}\label{eq:T_crit-big}
	\frac{k_B T_{\rm {crit}}}{\hbar \omega}
	\sim  \frac{\delta/\omega}{2 \ln(\delta/\omega)}.
\end{equation}

\medskip

\begin{figure} [t]
\centerline{\epsfig{file=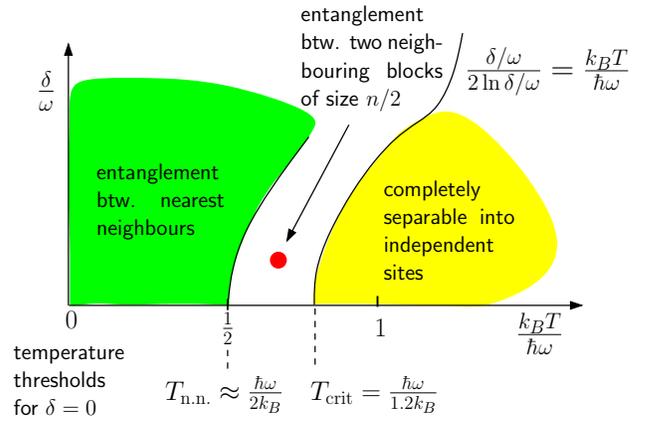, width=8.2cm}} 
\vspace*{13pt}
\caption{\label{fig:T-line} Schematic of how the entanglement 
  	is distributed between the spatial sites for varying $\delta$ ($n$ 
           is thought to be large). When $\delta =0$, bipartite entanglement 
           resides between nearest neighbours (green) for temperatures up to 
           $T_{\rm {n.n.}} = \frac{\hbar \omega}{2 \, k_B}$ and to a lesser degree 
           and for smaller temperatures between non-nearest neighbours~\cite{Janet}.
           On the other hand, according to 
           Eq.~(\ref{eq:T_crit-low}), above $T_{\rm {crit}} = \hbar \omega /1.2 \, k_B$
           there is no entanglement for any possible split (yellow). Between $T_{\rm {n.n.}}$ 
           and $T_{\rm {crit}}$ entanglement must be present but not simply 
           between two single sites. Audenaert \textit{et al.}~\cite{AEPW}
           have reported numerical results showing that entanglement above 
           $T_{\rm {n.n.}}$ exists between two neighbouring \emph{blocks} of size 
           $n/2$ up to a temperature of  $T_{\rm blocks} \approx 2.8\hbar \omega / \sqrt{20} \, k_B$ 
           for the case $\omega = \sqrt{20} \delta$ (the unitless value $2.8$ appears to
           be the critical value according to Figure 12 in~\cite{AEPW};
           our expression is obtained by retaining all units).
           When $\delta$ is large compared to $\omega$, the frequency spectrum Eq.~(\ref{eq:dispersion}) 
           shrinks to essentially one dominating frequency, $\omega_j \approx \delta$. 
           The mixing of frequencies occurring at increasing temperatures becomes 
           therefore irrelevant and entanglement persist up to higher temperatures, 
           the higher $\delta$ becomes. By Eq.~(\ref{eq:T_crit-big}), 
           the asymptotic behaviour is 
           $\frac{k_B T_{\rm {crit}}}{\hbar \omega} \sim \frac{\delta/\omega}{2\ln(\delta/\omega)}$.}
\end{figure}

It turns out that the order of magnitude of temperatures where entanglement can occur 
is indeed set already by the nearest-neighbour entanglement, with full separability kicking 
in at only roughly double the temperature where nearest neighbour (n.n.) entanglement
disappears. A qualitative picture of this behaviour in comparison to the nearest neighbour 
entanglement is shown in Fig.~\ref{fig:T-line}.

\section{Measuring the Entanglement at Sub-critical Temperature}
\label{sec:e-measure}
\noindent
Now that we know a pretty good bound on the entanglement-critical temperature for general systems, and indeed the exact value for translation-symmetric systems of the type (\ref{eq:kinetic+potential}), we proceed to an attempt to quantify the amount of entanglement below the critical temperature. In this section, we stick to systems of the latter type, even though one obtains bounds on the entanglement for more general systems in the same way as before. 

The most straightforward thing to do is to measure by how much we fail to satisfy the inequality $\gamma \geq \bigoplus_{j=1}^n \eta_j$, compare Eq.~(\ref{eq:multi-sep}). E.g.~one may set 
\begin{equation}  \label{eq:P-def}
  P(\gamma) := \max \bigl\{ 0\leq p\leq 1 : 
  	\gamma \geq \, p \, \bigoplus_{j=1}^n \, \eta_j
  	\quad  \mbox{for CMs} \quad \eta_j \bigr\},
\end{equation}
and consider $-\log P(\gamma)$ as a measure how far $\gamma$ is from being separable (see~\cite{GiedkeCirac}, where the above is considered for bipartite splits and denoted $V$, and~\cite{Anders-dipl}). In~\cite{GiedkeCirac} it is shown that this is indeed a \emph{Gaussian} entanglement monotone (i.e.~a measure monotonic under Gaussian LOCC).

As before, by translation symmetry and the absence of mixed terms on the covariance matrix, we may assume w.l.o.g.~that all $\eta_j$ in Eq.~(\ref{eq:P-def}) equal some 
\begin{equation}
  \eta_0 = \left[ \begin{array}{cc}
                      \frac{\hbar}{2m\omega_0} & 0 \\
                      0 & \frac{m\hbar\omega_0}{2}
                    \end{array} \right],
\end{equation}
yielding
\begin{equation}
  \label{eq:P-formula}
  P(\gamma) = \max_{\omega_0} \min_{j=1\ldots n}
                   \left\{ \frac{\omega_j}{\omega_0}\coth\frac{\beta\hbar\omega_j}{2},
                           \frac{\omega_0}{\omega_j}\coth\frac{\beta\hbar\omega_j}{2}, 1 \right\}.
\end{equation}
This is not as easy to evaluate as the critical temperature (where $P(\gamma)$ becomes $1$) but still remarkably simple.
Note that as before for the critical temperature, the value depends only on the frequency spectrum of $H$, not on the details of the phononic modes. Now, however, all the frequencies play a role in $P(\gamma)$ as a function of the inverse temperature $\beta$, not only the smallest and the largest.

\medskip
As a side remark, if we do the optimisation of Eq.~\eqref{eq:P-def} for a bipartite cut $A|B$ of the system, $P^{A|B}$ is related to other nice entanglement measures, specifically the Gaussian entanglement of formation $E_G^{A|B}(\gamma)$~\cite{Gaussian-EoF}, for covariance matrices $\gamma$. 
Namely, as shown there,
\begin{equation}
  E_G^{A|B}(\gamma) = \inf_{\gamma'} \left\{  E(\gamma') : \gamma \geq \gamma' \rm { and }
                                                                       \gamma' \rm { pure} \right\},
\end{equation} 
where $E(\gamma')$ is the entropy of entanglement of the pure entangled Gaussian state
with covariance matrix $\gamma'$.
Furthermore it is known~\cite{Gaussian-EoF} that a \emph{pure} Gaussian state $\gamma'$
is equivalent (via Gaussian local operations) to a direct sum of two-mode
squeezed states, $\gamma' = \bigoplus_k \gamma'(\tau_k)$, where
\begin{equation}
  \gamma'(\tau) = \left[\begin{array}{cccc}
                          \cosh \tau & 0           & \sinh \tau & 0 \\
                          0          & \cosh \tau  & 0          & -\sinh \tau \\
                          \sinh \tau & 0           & \cosh \tau & 0 \\
                          0          & -\sinh \tau & 0          & \cosh \tau
                        \end{array}\right].
\end{equation}
This means that, with $\tau_{\max}$ denoting the maximum of the $\tau_k$, $E(\gamma') = \sum_k E\bigl( \gamma'(\tau_k) \bigr) \geq E\bigl( \gamma'(\tau_{\max}) \bigr)$, while
\begin{equation}
	\begin{split}
		P^{A|B}(\gamma) \geq P^{A|B}(\gamma') = \min_k P^{A|B}\bigl( \gamma'(\tau_k) \bigr) \\
		= P^{A|B}\bigl( \gamma'(\tau_{\max}) \bigr)    = e^{-\tau_{\max}},
	\end{split}
\end{equation}
where the first inequality follows from $\gamma \geq \gamma'$ and the definition of $P$,
and the first and third equality sign are facts proven in~\cite{GiedkeCirac}.
Hence, $-\ln P^{A|B}(\gamma) \leq \tau_{\max}$.

Using the formula for the entanglement entropy of the two-mode squeezed state $\gamma'(\tau)$~\cite{Gaussian-EoF}, we arrive at the bound
\begin{equation}
  E_G^{A|B}(\gamma) \geq {\cal H}\bigl( -\ln P^{A|B}(\gamma) \bigr),
\end{equation}
with ${\cal H}(\tau) := \cosh^2 \tau \log \cosh^2 \tau - \sinh^2 \tau \log \sinh^2 \tau$,
a kind of ``hyperbolic entropy''. This results -- after some elementary transformation -- 
in a relation
\begin{equation}
  E_G^{A|B}(\gamma) \geq -2\log P^{A|B}(\gamma) + \Delta\bigl( P^{A|B}(\gamma) \bigr),
\end{equation}
where $\Delta$ is a monotonic and convex correction (in $P$) with the properties $\Delta(1)=0$, $\Delta(0)=\frac{1}{\ln 2}-2 = -0.557304\ldots$.

\section{Discussion and Conclusions} \label{sec:discussion}
\noindent
We have shown that in all systems of finitely many (harmonically) coupled harmonic oscillators, the thermal state at sufficiently high temperature is fully separable. We note, that our argument was quite independent of the actual Hamiltonian, and may still hold for a much wider range of systems.
For systems whose Hamiltonian has separate kinetic and potential terms for the physical oscillators in space we have found a \emph{bound} on the corresponding critical temperature, 
\begin{equation} \label{eq:energy-scale}
  	k_B T_{\rm {crit}} \leq \frac{\hbar \omega_{\max}}{\sigma(r)},
\end{equation}
where $\sigma(r)$ is a universal scaling function of the spectral ratio $r = \frac{\omega_{\max}}{\omega_{\min}}$ between the maximal (minimal) eigenfrequencies of the Hamiltonian $\omega_{\max}$ ($\omega_{\min}$). This is a quite intuitive result, as it says that the temperature has to be just high enough so that all the normal modes are participating sufficiently in the thermal state.
In general one can of course not expect the above formula to give the correct cutoff temperature for entanglement, as it neither refers to the full eigenfrequency spectrum, nor at all to the actual form of those eigenmodes themselves. For example, as we pointed out in Section~\ref{sec:normal}, a system of decoupled oscillators with eigenfrequencies $\omega_j$ is disentangled at \emph{all} temperatures.
For Hamiltonians with sufficient translation symmetry, however, such as periodic chains or lattices, the right hand side of Eq.~(\ref{eq:energy-scale}) turns out to give \emph{exactly}  the critical temperature. The intuition why this can happen is that the translation symmetry of the Hamiltonian sufficiently determines the entangled structure of the eigenmodes.

\medskip
For the periodic harmonic chain, our results made for some interesting 
comparison with earlier work, such as~\cite{AEPW} and~\cite{Janet}. There, only two 
parameters govern the behaviour of the system, the interaction strength $\omega$ 
and the on-site potential $\delta$, see Eq.~(\ref{eq:hamiltonian-ring}). We find a 
\emph{phase diagram} of different entanglement/separability regimes as shown in 
Fig.~\ref{fig:T-line}. The general case of quadratic Hamiltonian is qualitatively 
similar in terms of the maximal eigenfrequency $\omega_{\max}$ and the ratio 
$r = \frac{\omega_{\max}}{\omega_{\min}}$.

\medskip

Based on the Gaussian separability criterion, we were then able to even quantify the entanglement at sub-critical temperatures via a certain distance to the separable set. The most amazing aspect of the resulting formulae, Eqs.~(\ref{eq:P-formula}) and~(\ref{eq:energy-scale}), is that these expressions are given in terms of the normal mode spectrum alone. This is because the entanglement properties of the normal modes themselves are sufficiently captured by the translation symmetry. It should be noted, however, that the $P$-measure of entanglement, though related to the (Gaussian) entanglement of formation for bipartite cuts, is not an extensive quantity, hence it can not be used to extend the results of~\cite{CEPD-area} on entanglement-area laws for the ground state of harmonic systems.

\medskip
We had to leave open a number of intriguing questions, among them the calculation or at least estimation of an extensive entanglement measure, such as (Gaussian) entanglement of formation between two disjoint subsets of oscillators, especially the case of a region and its complement in a lattice, as a function of the temperature and the shape of the regions, for example like the area laws for ground states discussed in \cite{CEPD-area}.
One can also speculate about the existence of bound entanglement in the thermal state, more precisely, temperatures at which the state is PPT  (positive partial transpose) \emph{and}  entangled. Related to this is the more general question about the \emph{kind} of entanglement in the thermal state, in particular whether it can be characterised by bipartite entanglement or if it is genuinely multipartite. 
A recent paper \cite{Cavalcanti07} addresses exactly this issue and our results combined with those in \cite{Cavalcanti07} prove, that fully PPT entangled states, i.e. entangled states such that all bipartite partitions are PPT, cannot be obtained for the harmonic systems investigated here. This is because the critical temperatures for the negativity of the even-odd partition in  \cite{Cavalcanti07} coincide with our Eq.~\eqref{eq:beta-propto-spectrum}  for full separability. Thus, the corresponding thermal states become PPT and  fully separable for the same temperature.

\medskip
Another, deeper and physically more interesting, question is whether the regions in our \emph{entanglement phase diagram}, Fig.~\ref{fig:T-line}, have any actual interpretation as different phases of the system in the sense of physically radically different \emph{orders}. Clearly, we do not expect all the different boundaries to be phase transitions crossovers, just as we don't expect all possible entanglement measures to be thermodynamically relevant.

\medskip
For certain approaches to the quantum-to-classical transition stipulating its being connected to the loss of entanglement the following observation may be relevant. 
For a macroscopically large system, we would expect that the quantum description might have faster and faster normal modes, requiring that the temperature guaranteeing separability goes to infinity. 
This clearly means that for a reasonable emergence of classicality, full separability of the state is asking too much; on the other hand, it can be interpreted as saying that even large and hot systems retain some quantum character hidden in very high energy degrees of freedom. 

A different but related issue is connected to the observation that
indeed, the set of entangled states is dense in the set of all states for continuous variable systems \cite{Eisert02}. Hence one may ask how realistic the separable states found here are or whether they are artificial mathematical constructs. In other words, the actual state of the system may never be the exact thermal state, which would be separable above $T_{\rm{crit}}$, but an entangled non-equilibrium state nearby.  This question is identical to asking how clear a cut exists between a quantum and classical world. The paper \cite{Eisert02} has a very interesting point to make about this: namely, that as long as we only look at states of bounded energy (say, of the same scale as the thermal state under consideration) in the vicinity of the thermal state, then the amount of entanglement in
theses states is bounded by a function of the distance (measured via the trace norm). In other words, while we cannot be sure that the actual state, fluctuating a tiny bit away from the exact thermal state, is indeed separable, we \emph{do} know that the amount of entanglement in it must be very small -- as long as we can control its mean energy and distance from the thermal state.

\medskip
Finally, the choice of the `parties' to find entanglement between, which we addressed in Remark 1, becomes even more urgent for the continuum limit of the harmonic model. In the limit of infinitely many harmonic oscillators each site turns into a point in space  that can oscillate and the system becomes a quantum field. The entanglement can now exist between modes of the field  instead of single particles \cite{vanEnk03,vanEnk05} and one is left with the choice of which modes to talk about \cite{Zanardi01}. The decoupled, unentangled normal modes in the lattice, the lattice vibrations, become continuous single-particle energy-eigenfunctions and the entanglement between the former discrete sites in the harmonic chain becomes entanglement between points in space in the quantum field. In a rescaled and non-diverging theory the interaction between spatial points, $\omega$, is infinitesimally small and hence no entanglement can be found between them. However, analogous to the block entanglement in finite harmonic chains, discussed in \cite{AEPW}, the entanglement between finite volumes in space can be tested as done in \cite{Heaney07,HAV,Janet-etal}.  

\medskip

\noindent {\bf Acknowledgements.} 
JA is supported by the Gottlieb Daimler und Karl Benz Stiftung.
AW acknowledges the support of the U.K.~EPSRC through an Advanced
Research Fellowship and through the ``IRC QIP''; furthermore support
by the European Commission via project ``QAP'' (contract-no.~IST-2005-15848).

\vfill


\begin{thebibliography}{000}
\bibitem{footnote1} The symplectic eigenvalues of a covariance matrix $\gamma$ are simply the eigenvalues that are preserved under symplectic transformations and therefore condense the essential information of the covariance matrix, see for instance \cite{Anders-dipl} or references therein. They can be calculated as the absolute values of the common eigenvalues of the matrix $i \sigma \gamma$. Symplectic eigenvalues always appear with double multiplicity since the position and the momentum are treated interchangeably. This implies that for a single-mode covariance matrix ($2 \times 2$), its determinant is just the square of the single symplectic eigenvalue.
\end{thebibliography}
\end{document}